\documentclass[
cleveref
]{iacrtrans}
\title{Improved phase noise measurement for multi-ring oscillator based TRNG }

\author{David Lubicz \and Maciej Skorski}

\usepackage{easy-todo}
\usepackage[norelsize,linesnumbered,algoruled]{algorithm2e} 
\usepackage{listings}
\usepackage{pgfplots}
\usepackage{svg}
\usepgfplotslibrary{groupplots}
\usepackage[frozencache=true,cachedir=minted-cache]{minted} 
\graphicspath{{figures/}}
\usepackage{mdframed}
\usepackage{easy-todo}
\pgfplotsset{compat=1.18}
\usepackage{tikz}
\usetikzlibrary{positioning,shapes,arrows, arrows.meta}
\usetikzlibrary{external}
\usepackage{subcaption}

\usepackage{nicematrix}

\usepackage{hyperref}
\hypersetup{
    colorlinks = false,
    citecolor = red,
}
\usepackage[nameinlink]{cleveref}

\newcommand\getrand{\stackrel{R}{\leftarrow}}

\author{D. Lubicz \inst{1}, M. Skorski \inst{2}} 
\institute{
    Institut de Recherche Mathematiques de Rennes,
    263 avenue du Général Leclerc, CS 74205
35042 RENNES Cédex
\and 
Université Jean Monnet Saint-Etienne, CNRS, Institut d Optique Graduate School, Laboratoire Hubert Curien UMR 5516, F-42023, SAINT-ETIENNE, France
}
\keywords{Ring oscillators, True Random Number Generators} 

\newmdtheoremenv{framedclaim}{Claim}

\begin{document}

\maketitle

\begin{abstract}
The aim of this paper is to describe a way to improve the reliability of the measurement of the statistical parameters of the phase noise in a multi-ring oscillator-based TRNG. This is necessary to guarantee that the entropy rate is within the bounds prescribed by standards or security specifications. According to the literature, to filter out global noises which may strongly affect the measurement of the phase noise parameters, it is necessary to perform a differential measure. But a differential measurement only returns the parameters of the phase noise resulting of the composition of the noises of two oscillators whereas jitters parameters of individual oscillators are required to compute the entropy rate of a multi-ring oscillator-based TRNG. In this paper, we revisit the "jitter transfer principle" in conjunction with a tweaked design of an oscillator based TRNG to enjoy the precision of differential measures and, at the same time, obtain jitter parameters of individual oscillators. We show the relevance of our method with simulations and experiments with hardware implementations.
\end{abstract}


\section{Introduction}\label{sec:intro}
Precise entropy evaluation, which is required by security standards such as AIS 31 \cite{KS11}, is essential for guaranteeing unpredictability of generated number and thus security.
To specify the entropy rate at the TRNG output, the designer has to:
\begin{enumerate}
\item  propose a parameterized stochastic model or use some existing model appearing in the literature, such as [BLMT11], which specifies distribution of RNG output values depending on some non-manipulable physical sources of randomness, e.g. the thermal noise ;
\item  measure (preferably online and inside the device) input parameters of the model;
\item  use the model to compute the entropy rate.
\end{enumerate}
In practice, the designer has to determine a trade-off between the bit rate and entropy rate: the higher the throughput is, the lower will be the entropy rate. Fortunately, in the case of RNGs using free running oscillators, both bit rate and entropy rate can be increased by increasing the number of oscillators. But this is done at the expense of a more complex implementation and stochastic model, which needs rigorous estimation of contribution of each oscillator on the total entropy rate.

The jitter is a complex phenomenon resulting from different kinds of physical noises \cite{DLNO}. While both the global noises and local noises determine the clock jitter, the former do not contribute to the entropy rate of the TRNG because they can be manipulated or computed by an attacker. Consequently, if the global noises are not excluded from entropy estimation, the obtained entropy rate can be significantly overestimated. It was clearly demonstrated in \cite{BBFV} that the differential jitter measurement reduces significantly the contribution of global noises on the output entropy rate. Several embedded jitter measurement methods have already been published and evaluated \cite{DBLP:conf/cardis/GarayBFHM22}, while all of them are based on the differential principle. However, the differential measure will always output the parameters of the jitter resulting in the combination of contribution of two oscillators.
However, in the TRNGs using multiple freely running oscillators \cite{MR4712007}, in order to avoid entropy overestimation based on the model, the contribution of each oscillator has to be considered separately and not in couples of rings used to measure the jitter. Up to now, no solution has been proposed for this serious problem. Our main contribution is thus a way to obtain individual jitters of oscillators by performing differential measurements.

Our key tool is the quantitative description of edge events in  oscillators paired for differential measurement, based on technical development in the phase domain. We would like to highlight that studying edge events in "paired" oscillators needs the phase domain and cannot be done with a discrete model of jittered edge arrivals. Our paper shows how these two are related and that certain intuitions developed in time-domain are (nearly) matching the dynamic in the phase domain.
In regard to electronic principles, our phase-domain model encapsulates various hardware-dependent constants in the phase volatility, yet captures only the thermal noise contribution. Addressing other noises could increase the entropy but remains an open problem and a subject of intensive research.

The paper is organised as follows: 
in Section \ref{sec:principle}, we introduce the main technical ingredient and results of the paper which is the jitter's transfer principle and its error bounds. Then in \Cref{sec:applications}, we explain how to apply these results to the computation of the entropy rate of an oscillator based TRNG. In \Cref{sec:impl}, we describe an hardware implementation and experiments to demonstrate the effectiveness of our methods. Finally, in \Cref{sec:conclusion}, we summarise the main conclusions of our work. We have gathered in \Cref{sec:proofs} all the proofs of our results.
The supplementary material (Python code) is  shared anonymously on the OSF platform via the link 
\url{https://osf.io/zcj92/?view_only=ded350214f1349b2919e5ef22471c67a}.

\section{The jitter's transfer principle}\label{sec:principle}

\subsection{Preliminaries}
Consider two oscillators $O_0$, $O_1$ producing the signals $s_0,s_1$ defined by
\begin{align}\label{eq:model}
s_0(t) &= w(\phi_0 + f_0 t+\xi^{0}_t ), \\ 
s_1(t) & = w(\phi_1 + f_1 t+\xi^{1}_t ),
\end{align}
where $w$ is a 1-periodic square wave, $\phi_i$ are initial phase locations, $f_i$ are frequencies and $\xi^{i}_t$ are (stochastic) phase noises modelled by using Wiener processes with volatility $\sigma_i$. The units of measure are, respectively, $\mathrm{Hz}=s^{-1}$ for $f_i$ and $s^{-\frac{1}{2}}$ for $\sigma_i$~\footnote{Consistently with the standard deviation of $\xi^{i}_t$ which equals $\sqrt{t}\sigma_i$.}.

\begin{figure}
\includesvg[inkscapelatex=false,width=0.99\linewidth]{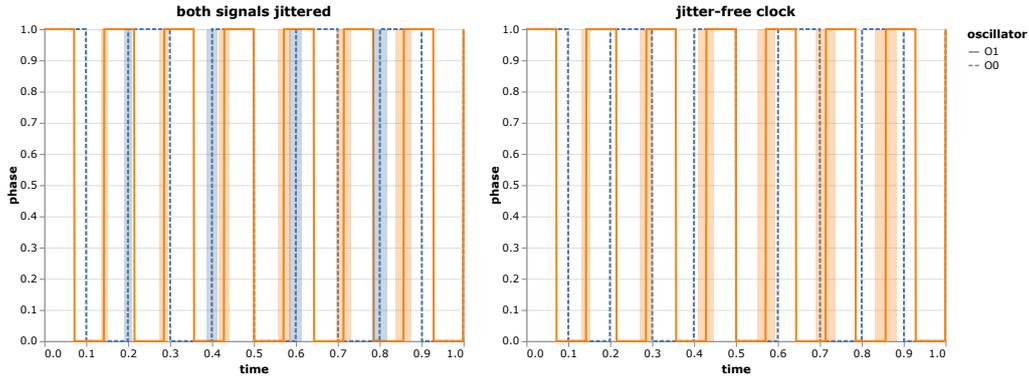}
\caption{Jitter Transfer Principle: both signals jittered (left) versus jitter transferred (right). The  oscillators in this toy example have frequencies $f_0 = 5,f_1=7$ and volatilities $\sigma_0=\sigma_1 = 0.02$ (seen in standard error bands around rising edges).
}
\label{fig:jitter_transfer}
\end{figure}

In the classical design of an elementary oscillator-based TRNG (EO-TRNG) the signal of $O_1$ is sampled by that of $O_0$ using a type-D flip-flop. Thus, the bits are generated by sampling $s_1$ at the rising edges of the signal $s_0$, which can be defined formally as
\begin{align}\label{eq:bit_model}
    b_k = s_1(T_k), \quad T_k \triangleq \inf\{t: \phi_0 +f_0 t + \xi^{0}_t = k\},\quad k=1,2,\ldots.
\end{align}
To simplify stochastic analysis, we would like to prove that the system of two oscillators is equivalent to a one with jitter of $s_0$ “fully transferred” to $s_1$, as stated informally below: 
\begin{framedclaim}[Jitter Transfer Principle]
The bit sequence $(b_k)$ generated by the system of two noisy oscillators in \eqref{eq:model} is approximately indistinguishable from the sequence generated by a system in which $O_1$ encapsulates all noise, while $O_0$ is noise-free.
\end{framedclaim}
The principle is illustrated in \Cref{fig:jitter_transfer}.

One version of this jitter transfer principle was described and proved in \cite{baudet_security_2011}. In that paper, the authors describe a statistical model aimed at computing the entropy rate of an EO-TRNG. However one shortcoming of the statistical model is that it relies on the simplifying hypothesis that the sampling signal $s_0$ is stable.
In order to take into account the case that the sampling signal is itself a jittery clock signal, the jitter transfer principle is used to translate the output of the measurements into the parameters of the statistical model used to compute the entropy rate.

The statistical equivalence stated by the jitter transfer principle is just an approximation and depends on the hypothesis that the volatility of the jitter is small compared to its period. One shortcoming of the jitter transfer principle of \cite{baudet_security_2011} is that it does not provide the error bounds needed to certify the computation of the entropy rate of the TRNG. Moreover, the jitter phenomenon can be defined in different ways including as a distribution of events such as the rising edges of a signal in the time domain or in the phase domain such as in (\ref{eq:bit_model}). In practice, the jitter is either measured (for
instance with an oscilloscope) as a distribution on the rising edges of a clock signal in the
time domain or using the internal measurement method in the phase domain. Thus the jitter
transfer principle is mostly used between two distributions in the time domain or
between two distributions in the phase domain. But in \cite{baudet_security_2011}, the jitter transfer principle applies between one distribution in the time domain and another in the phase domain meaning that in most cases, it is not applicable. 

The next section provides a simpler proof of jitter transfer principle~\cite{baudet_security_2011} that rcomes with closed formulas for error bounds.  

Before presenting technical results, we first need to introduce some probability laws.
Recall that a \emph{normal variance–mean mixture}~\cite{barndorff-nielsen_normal_1982} is a distribution of the form
$Y\sim \alpha + \beta V+\sigma \sqrt{V}\cdot \mathsf{N}(0,1)$,
where $\alpha,\beta,\sigma$ are real constants with $\sigma>0$, and $V$ is a non-negative continuous random variable. The \emph{inverse Gaussian distribution} $\mathsf{IG}(\mu,\lambda)$ has density $f(x;\mu,\lambda) = \sqrt\frac{\lambda}{2 \pi x^3} \exp\biggl(-\frac{\lambda (x-\mu)^2}{2 \mu^2 x}\biggr)$. The normal-inverse Gaussian distribution $\mathsf{NIG}(\alpha,\beta,\mu,\delta)$ has the density
$f(x; \alpha, \beta, \mu, \delta) =
        \frac{\alpha\delta K_1\left(\alpha\sqrt{\delta^2 + (x - \mu)^2}\right)}
        {\pi \sqrt{\delta^2 + (x - \mu)^2}} \,
        e^{\delta \sqrt{\alpha^2 - \beta^2} + \beta (x - \mu)}$ where $K_1$ is the modified second-kind Bessel function of order 1~\cite{abramowitzHandbookMathematicalFunctions1974}.

\subsection{Analytical Distributions of Rising Edges and Sampled Phases}

\Cref{thm:main} below addresses the challenge of characterizing precisely (not relying on approximation) the distribution of the phase position of the sampled oscillator.

\begin{theorem}\label{thm:main}

The phase of the sampled signal at the $k$-th rising edge of the clock signal is distributed as the normal variance-mean mixture
\begin{align}\label{eq:mean_var_mix}
\Phi_1(T_k) \sim \mathsf{N}(\phi_1+f_1T_k,\sigma_1^2 T_k),
\end{align}
where the mixing distribution $T_k$ is the time it takes to reach the $k$-th rising edge of $s_0$ and is given by the inverse normal distribution
\begin{align}
\begin{aligned}
T_k & \sim \mathsf{IG}\left(\frac{k-\phi_0}{f_0},\frac{(k-\phi_0)^2}{\sigma_0^2}\right),
\end{aligned}
\end{align}
and the $k$-th clock period is distributed as 
\begin{align}
    T_{k+1}-T_{k} \sim \mathsf{IG}\left(\frac{1}{f_0},\frac{1}{\sigma_0^2}\right).
\end{align}
The change in the sampled oscillator phase during one clock period is given by the variance-mean mixture
\begin{align}
\Phi_1(T_{k+1})-\Phi_1(T_{k}) & \sim \mathsf{N}(f_1 Z, \sigma_1^2 Z),\quad  Z  \sim \mathsf{IG}\left(\frac{1}{f_0},\frac{1}{\sigma_0^2}\right),
\end{align}
which, in terms of the normal-inverse Gaussian distribution, can be written as
\begin{align}
  \Phi_1(T_{k+1})-\Phi_1(T_{k})  \sim \mathsf{NIG}\left(
    \sqrt{\frac{f_{0}^{2}}{\sigma_{0}^{2} \sigma_{1}^{2}} + \frac{f_{1}^{2}}{\sigma_{1}^{4}}},
    \frac{f_{1}}{\sigma_{1}^{2}},
    0,
    \frac{\sigma_{1}}{\sigma_{0}}
  \right).
\end{align}
\end{theorem}

\begin{corollary}\label{cor:mean_variance}
The mean and variance of the target signal phase at $T_k$ equal
\begin{align}
\begin{aligned}
    \mathbf{E}[\Phi_1(T_k)] & = \phi_1 + f_1 \mathbf{E}[T_k] =  \phi_1 + (k-\phi_0)\cdot\frac{f_1}{f_0}, \\
    \mathbf{Var}[\Phi_1(T_k)] & = f_1^2 \mathbf{Var}[T_k] + \sigma_1^2 \mathbf{E}[T_k] = \frac{f_1^2}{f_0^3}\cdot (k-\phi_0) \cdot \sigma_0^2 + (k-\phi_0)\cdot \frac{\sigma_1^2}{f_0}.
\end{aligned}
\end{align}
\end{corollary}

The following corollary relates the volatility to the jitter observed over time:
\begin{corollary}[Relating Phase Noise and Time Jitter]\label{cor:relating_jitters}
Let $T\triangleq T_{k+1}-T_k$ be the duration of the $k$-th clock cycle. Then 
\begin{align}
\frac{{\mathbf{Var}[T]}}{\mathbf{E}[T]^2} =  {\frac{\sigma^2}{f}}.
\end{align}
\end{corollary}
\begin{example}
Consider a jittered oscillator with the cycle duration $T$ such that $\mathbf{Var}[T] = 0.094(\mathrm{ns})^2$ and $\frac{1}{f_1}=\mathbf{E}[T]=33.4\mathrm{ns}$. During one cycle of jitter-free clock with ${f_0}=50\mathrm{MHz}$,
the phase variance would increase by $Q=\frac{\sigma_1^2}{f_0} = \frac{{\mathbf{Var}[T]}}{\mathbf{E}[T]^2} \cdot \frac{f_1}{f_0} = \frac{0.094(\mathrm{ns})^2 }{ (33.4 \mathrm{ns})^3 50\mathrm{MHz}}\approx \frac{1}{198190}$ (compare with the nearly matching approximation in Section 4.1 in \cite{baudet_security_2011}). 
\end{example}

\subsection{Normal Approximation}

Although the normal-inverse Gaussian distribution has analytical formulas and is supported by statistical software, in certain regimes - and this turns out to be the "small jitter" case - it can be well approximated by a normal distribution. We now prove this fact formally.

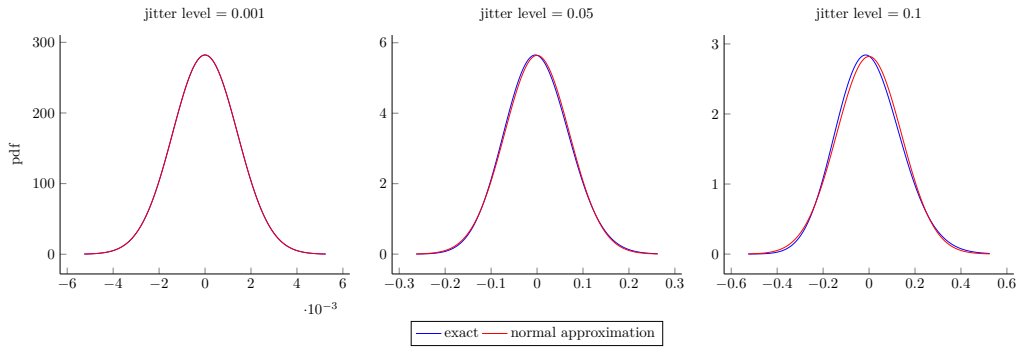
\begin{figure}[h!]
\resizebox{0.99\textwidth}{!}{
\begin{tikzpicture}
    \begin{groupplot}[group style={
                      group name=myplot,
                      group size= 3 by 1},
                      axis y line*=left,
                      axis x line*=bottom]
    \nextgroupplot[title={jitter level $=0.001$}, ylabel={pdf}]
    \pgfplotstableread[col sep=comma]{./data/discrepancy_0.001.csv}\datatable
    \addplot[color=blue] table[x=phase, y=pdfexact, col sep=comma]{\datatable};
    \addplot[color=red] table[x=phase, y=pdfapprox, col sep=comma]{\datatable};
    \coordinate (first) at (rel axis cs:0,0);
    \nextgroupplot[title={jitter level $=0.05$},ylabel={}]
    \pgfplotstableread[col sep=comma]{./data/discrepancy_0.050.csv}\datatable
    \addplot[color=blue] table[x=phase, y=pdfexact, col sep=comma]{\datatable};
    \addplot[color=red] table[x=phase, y=pdfapprox, col sep=comma]{\datatable};
    \nextgroupplot[title={jitter level $=0.1$},ylabel={},
    legend style={at={($(0,0)+(1cm,1cm)$)},
    legend columns=3,fill=none,draw=black,anchor=center,align=center},
    legend to name={fred}
    ]
    \pgfplotstableread[col sep=comma]{./data/discrepancy_0.100.csv}\datatable
    \addplot[color=blue] table[x=phase, y=pdfexact, col sep=comma]{\datatable};
    \addlegendentry{exact};   
    \addplot[color=red] table[x=phase, y=pdfapprox, col sep=comma]{\datatable};
    \coordinate (last) at (rel axis cs:1,0);
    \addlegendentry{normal approximation};
\end{groupplot}
\path (first.south) -- (last.south) node [midway, below={1cm of last.south}] {\pgfplotslegendfromname{fred}};
\end{tikzpicture}
}
\caption{Quality of the normal approximation in the Jitter Transfer Principle (global view). This example assumes similar frequencies and volatilities: $f_1\approx f_0$, $\sigma_1\approx \sigma_0$, and studies various jitter levels defined as $\sigma[T_i]/T_i = \sqrt{\sigma_i^2/f_i}$.}
\label{fig:approx}
\end{figure}
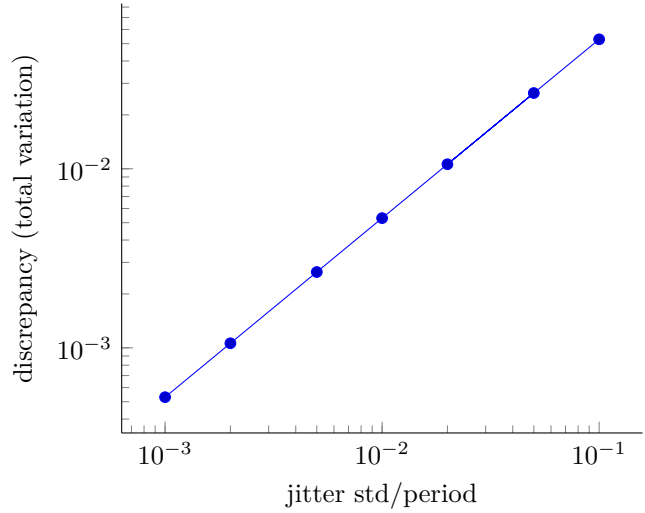
\begin{figure}
\centering
\begin{tikzpicture}
\begin{axis}[
    xmode=log,
    ymode=log,
    axis y line*=left,
    axis x line*=bottom,
    xlabel = {jitter std/period},
    ylabel = {discrepancy (total variation)}
]
\pgfplotstableread[col sep=comma]{./data/jitter_discrepancy.csv}\datatable
    \addplot+[] table[x=jitter, y=discrepancy, col sep=comma]{\datatable}; 
\end{axis}
\end{tikzpicture}
\caption{Quality of the normal approximation in the Jitter Transfer Principle (discrepancy). This example assumes similar frequencies and volatilities: $f_1\approx f_0$, $\sigma_1\approx \sigma_0$, and studies various jitter levels defined as $\sigma[T_i]/T_i = \sqrt{\sigma_i^2/f_i}$.}
\label{fig:approx_regression}
\end{figure}

\begin{theorem}\label{thm:normal_approx}
Let $\mu=\frac{f_1}{f_0}$ and $v=\sqrt{\frac{\sigma_{1}^{2}}{f_{0}} + \frac{f_{1}^{2} \sigma_{0}^{2}}{f_{0}^{3}}}$. Then, provided that
\begin{align}
    \frac{\sigma_0^2}{f_0} \to 0,
\end{align}
we have the convergence in distribution
\begin{align}
\frac{\phi_1(T_{k+1})-\phi_1(T_k)-\mu}{v} \overset{d}{\longrightarrow} \mathsf{N}(0,1),
\end{align}
which holds uniformly in $\sigma_1,f_1$.
\end{theorem}
The quality of normal approximation can be evaluated using the \emph{explicit}
normal-inverse Gaussian distribution given in \Cref{thm:main}. See \Cref{fig:approx} and
\Cref{fig:approx_regression}
below for a comparison.

\begin{remark}
Explicit bounds can be obtained by quantifying the distance between characteristic functions of the normal-inverse Gaussian and Gaussian distributions. See \cite{bobkov2016proximity} for a survey of related methods. Note that \Cref{fig:approx_regression} suggests that the total variation to the normal distribution grows as $O(\sigma_0/\sqrt{f_0})$.
\end{remark}

\subsection{Main results}

From the normal approximation in \Cref{thm:normal_approx} we obtain the following important fact
\begin{corollary}[Jitter Transfer Principle \cite{baudet_security_2011}]\label{cor:transfer}
Under the assumption
\begin{align}
    \sigma_0^2 \ll f_0,
\end{align}
sampling bits according to \Cref{eq:bit_model} is equivalent to sampling from a pair of oscillators where the clock $s_0$ is jitter-free and $s_1$ has volatility
\begin{align}
    \sigma = \sqrt{\frac{f_1^2}{f_0^2}\sigma_0^2+\sigma_1^2}.
\end{align}
\end{corollary}

\begin{remark}[Period Jitter Transfer]
Some authors~\cite{fischer_enhancing_2023} use the term \emph{period jitter} Onto refer to the variance of the (jittered) clock period. By \Cref{cor:relating_jitters}, the mapping from period jitter to phase noise is given by:
\begin{align*}
\begin{aligned}
\sigma_0^2 & \gets \sigma_0^2 f_0^3 \\
\sigma_1^2 & \gets \sigma_1^2 f_1^3 \\
{\sigma'}^2& \gets {\sigma'}^2 f_1^3,
\end{aligned}
\end{align*}
and \Cref{cor:transfer} expressed in period jitter reads as
\begin{align}
    \sigma' = \sqrt{\frac{f_0}{f_1}\sigma_0^2+\sigma_1^2}.
\end{align}
\end{remark}

\begin{remark}[Time Jitter Transfer]
Some works model the \emph{time jitter} (measured in seconds) instead of the phase noise, by considering the equation
$\phi_i(t)= \phi_0 + f_i \cdot (t+\xi^{i}_t ))$ (here the jitter term directly impacts the time, rather than the phase). This corresponds to substituting $\sigma_i \leftarrow f_i\cdot \sigma_i$ in \Cref{cor:transfer}, so that the formula becomes
\begin{align*}
\sigma' = \sqrt{\sigma_0^2+\sigma_1^2}.
\end{align*}

\end{remark}

\section{Applications}\label{sec:applications}

\subsection{Finding Oscillators' Volatilities}

\paragraph{Motivation}
In order to obtain a high entropy rate, improve reliability, account for possible flaws, ageing and malfunctions, an oscillator based TRNG is usually made of several EO-TRNGs the outputs of which are combined with a XOR. This structure is called multiring oscillator-based TRNG (MO-TRNG). In their paper \cite{MR4712007}, the authors provide an algorithm to compute the entropy rate of a MO-TRNG from the knowledge of the volatility of the phase noise of each of the rings it is made of. Note that the volatility of the phase noise manifests itself by a time shift in observable events, such as at the rising edge of a signal. In order to measure the jitter precisely and filter out non-random global noises, it is necessary to make a differential measurement \cite{baudet_security_2011}. By differential measurement, we mean that we measure the time shift due to jitter of a ring oscillator with respect to the time reference given by another ring oscillator. The output of such a differential measurement is the volatility of the composed jitter of a couple of oscillators. We explain how to use the jitter transfer principle on several couples of oscillators to \emph{recover the volatility of each individual oscillator} and hence enable the application of the method of \cite{MR4712007}.


\paragraph{Principle}
We illustrate exactly how our approach works combined with the internal measurement method of \cite{FL14} but it can be easily adapted to any kind of differential measurement, for instance~\cite{bernardLowCostPrecise2023}. We consider the MO-TRNG made of $O_i$, $i=0, \ldots, n$ ring oscillators with output signals $s_i=w(\phi_i + f_i t + \xi^i_t)$ where $\phi_i$ is the initial phase, $f_i$ is the mean frequency and $\xi^i_t$ is the Wiener process with drift $0$ and volatility $\sigma_i$. The signal of the ring oscillator $O_0$ samples the output signal of each of $O_i$, $i=1, \ldots, n$ by means of a type-D flip-flop. From the knowledge of all the $f_i$ and $\sigma_i$, using \cite{MR4712007}, we can compute a lower bound of the entropy rate of the MO-TRNG.

The method of \cite{FL14} makes it possible to compute the total jitter of a couple of oscillators $(O_i, O_j)$ by using some special statistical tests on the output bits of the EO-TRNG consisting of $O_i$ sampling the output signal of $O_j$. We assume that the hypothesis of \Cref{cor:transfer}
is fulfilled that is $\sigma_i^2 \ll f_i$. Then by  \Cref{cor:transfer}, the couple of oscillators $(O_i, O_j)$ is equivalent to a couple of oscillators $(O'_i, O'_j)$ where $O'_i$ is perfectly stable and the volatility of the phase noise of $O'_j$ is given by 
${\sigma'}_{i,j}^2 = \frac{f_j^2}{f_i^2} \sigma_i^2 + \sigma_j^2$. By performing an internal measurement for each possible $(O_i, O_j)$, we obtain ${\sigma'}^2_{i,j}$ and \Cref{cor:transfer} explains the connection between the output of the measurement and the individual jitter volatility we are looking for.

We now explain how we can recover the $\sigma_i$ from knowledge of the $\sigma'_{i,j}$. The first way to do that relies on the hypothesis that $\sigma_i^2 f_i$ is constant depending only on the technology and in particular independent of $i$. This hypothesis is supported by the intuition that if the electric signal goes through an environment with globally homogeneous properties then the variance in the thermal jitter volatility should be proportional to the length of the circuit and hence to the period of the ring oscillator. If this hypothesis is satisfied then we obtain that $\sigma_0^2 =\sigma_i^2 f_i/f_0$ so that 
\begin{equation}
{\sigma'}_{0,i}^2= \sigma_i^2 (1 + \frac{f_i^3}{f_0^3}).
\end{equation}
This last equation allows us to recover $\sigma_i$ from our knowledge of $\sigma'_{0,i}$. Actually, Fact 2 of \cite{FL14} makes it possible to recover the coefficients $\frac{f_i}{f_0} \mod 1$ and knowledge of the number of delay elements of each ring oscillator makes it possible to remove the ambiguity resulting from modulo $1$ so that we can recover
$\frac{f_i}{f_0}$ which are needed to carry out the computations.

\begin{figure}
    \centering
    \includegraphics[width=0.6\linewidth]{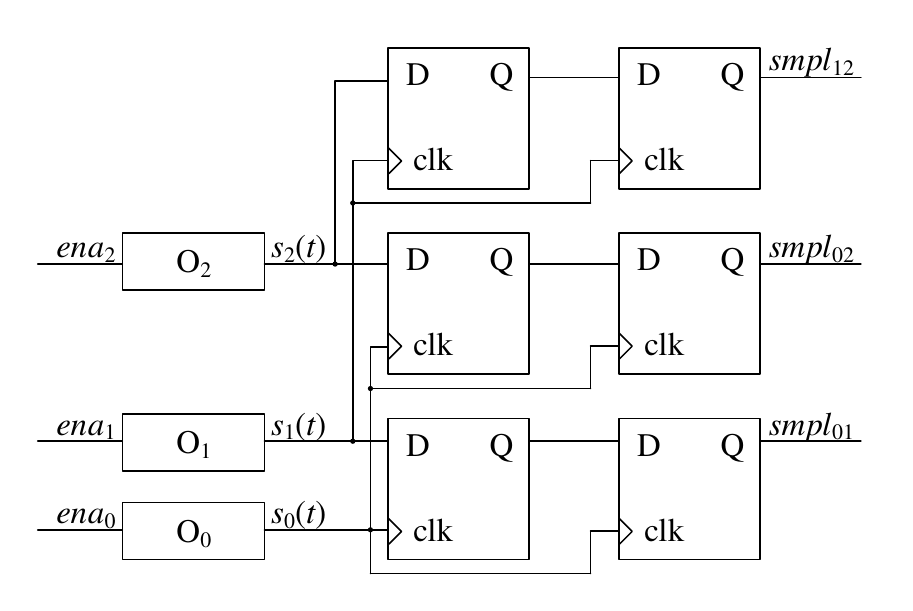}
    \caption{Schema of a MO-TRNG that allows to measure individual jitters}
    \label{fig:indms}
\end{figure}

If the hypothesis upon which the previous method relies is not fulfilled, we propose an alternative approach to get rid of the hypothesis although it requires a little more complex implementation. The idea behind the second method can be easily explained in the case of a MO-TRNG composed of the oscillators $O_1$ and $O_2$ whose signals are sampled by the signal of $O_0$ using a type-D flip-flop. As can be seen in \Cref{fig:indms}, we added a third flip-flop so that the signal of $O_2$ can be sampled using that of $O_1$. Hence compared with the simplest version of a MO-TRNG of \cite{MR4712007}, we only need to implement one more flip-flop as shown in Figure \ref{fig:indms} (it is also possible to use a multiplexer).

By performing the internal measurement of the couple of oscillators $(O_0, O_1)$, $(O_0, O_2)$ and $(O_1, O_2)$, we obtain respectively:
\begin{align}
\begin{aligned}
{\sigma'}_{0,1}^2&= \left(\frac{f_1}{f_0}\right)^2\sigma_0^2 + \sigma_1^2 \\
{\sigma'}_{0,2}^2&= \left(\frac{f_1}{f_0}\right)^2\sigma_2^2 + \sigma_{2}^2\\
{\sigma'}_{1,2}^2&= \left(\frac{f_2}{f_1}\right)^2\sigma_1^2 + \sigma_{2}^2.
\end{aligned}
\end{align}
It is easy to see that the preceding linear system in the unknown $\sigma_i^2$ always has a single solution and is well conditioned provided that the $f_i$ have the same magnitude. 

\paragraph{Verification for 3-ROs}

We checked that our method works properly when combined with the internal measurement method of \cite{FL14} in the case of a MO-TRNG with $3$ oscillators. Operating an EO-TRNG made of a couple $(O_0, O_1)$ of jittery ring oscillators where the signal of $O_0$ is used to sample the output of $O_1$ can be simulated with Algorithm \ref{algo:simul}. In this algorithm, in order to generate the sampling times, we can either draw samples following the real inverse Gaussian distribution according to the algorithm of \cite{wikiinverse} or use the normal approximation of the inverse Gaussian distribution provided by Theorem \ref{thm:normal_approx}. The former is more efficient but relies on the hypothesis that $\sigma_0^2 \ll f_0$. We checked that the two implementations produce results compatible with the error bounds of the normal approximation. 

\begin{algorithm}[h!]
\SetKwInOut{Input}{input}\SetKwInOut{Output}{output}

\SetKwComment{Comment}{/* Comment }{ */}
\Input{
    \begin{itemize}
	\item $(f_i, \sigma_i)$ the respective frequency and jitter's volatility of $O_0$ (sampling signal) and  $O_1$ (sampled signal) ;
 \item $\alpha$ the duty cycle of $O_1$;
	\item $n$ a number of output bits.
    \end{itemize}
}
\Output{
    $(b_j)_{j=1, \ldots, n}$ output bits of the EO-TRNG.
}
    \BlankLine
    $\phi \gets 0$ \tcc*{Initial phase of $O_1$} 
\For{$i =1$ to $n$} 
    {
    $P_{O_0} \getrand \mathsf{IG}(T_0, 1/\sigma_0)$  \tcc*{$\getrand$ means drawn following the right hand distribution
    \\ Replace line 3 by 
 $P_{O_0} \getrand \mathsf{N}(T_0, \sigma_0 T_0 \sqrt{T_0})$ to use the normal approximation of the sampling distribution}
 $\phi \getrand \mathsf{N}(\phi + \frac{1}{T_1}\, P_{O_0}, \sqrt{{\sigma}_1^2 \, P_{O_0}/T_1})$\;
 $\phi \gets \phi \mod 1$ \;
	\eIf{$\phi < \alpha$}
	{
	    $b_i \gets 1$\;
	    }
	    { $b_i \gets 0$\;
	    }
	}
	\Return $(b_i)_{i=1, \ldots, n}$\;
    \caption{Algorithm to simulate a EO-TRNG made of two jittery ring oscillators $O_0, O_1$ using the normal approximation of the inverse Gaussian distribution.}
\label{algo:simul}
\end{algorithm}

\begin{example}
We consider the EO-TRNG, which is made of three ring oscillators whose respective period and jitter volatility are listed in Table \ref{tab:tab1}. We assume that their duty cycle is $0.5$.

\begin{center}
\begin{table}
\begin{tabular}{|c|c|c|c|}
\cline{2-4}
\multicolumn{1}{c|}{}  & $O_0$ & $O_1$ & $O_2$ \\ \hline
Period & $T_0=1\, ms$ & $T_1=0.724\, ms$ & $T_2=0.652\, ms$ \\ \hline
jitter per $T_0$  & $\sigma_0(T_0)=1.00\, 10^{-3}$ & $\sigma_1(T_0)=2.00\, 10^{-3}$ & $\sigma_2(T_0)=3.00\, 10^{-3}$ \\ \hline
Measured jitter  & $\hat{\sigma}_0(T_0)= 0.96\, 10^{-3}$ &   
$\hat{\sigma}_1(T_0)=1.99\, 10^{-3}$ & $\hat{\sigma}_2(T_0)=3.00\, 10^{-3}$ \\ \hline
\end{tabular}
\caption{Volatility and measured volatility per $T_0$, here $\sigma_i(T_j)=\sqrt{T_j \sigma_i^2}$}\label{tab:tab1}
\end{table}
\end{center}

We denote by ${\sigma'}_{i,j}= \sqrt{\frac{T_i^2}{T_j^2}\sigma^2_i +\sigma^2_j}$ the total jitter volatility of the pair of oscillators  $(O_i, O_j)$ where $O_i$ is the sampling oscillator. We have to pay particular attention to the fact that the measurement method \cite{FL14} applied to the couple $(O_i, O_j)$ returns an approximation of the total jitter variance accumulated over a period of the sampling oscillator, that is $T_i {\sigma}^{\prime 2}_{i,j}$. It is then possible to recover ${\sigma}^{\prime 2}_{i,j}$ by dividing out by $T_i$. But the measurement of $T_i$ is not precise (because it is affected by deterministic global jitters) or not easily available inside the circuit.

In order to make our computations compliant with the outcome of easily available measurements techniques, we remark that what is actually needed to compute the entropy rate of a MO-TRNG using \cite[Algorithm 4]{fischer_enhancing_2023} is the variance of the jitter accumulated during one period of the sampling oscillator $O_0$. If we put 
$\sigma_i(T_j)=\sqrt{T_j \sigma_i^2}$ and 
$\sigma'_{i,j}(T_k)=\sqrt{T_k \sigma^{\prime 2}_{i,j}}$, using \Cref{cor:transfer}, we get:
\begin{align}
  \frac{T_i^3}{T_j^2 T_0} \sigma_i^2(T_0) +  \frac{T_i}{T_0} \sigma_j^2(T_0) =  {\hat{\sigma'}}_{i,j}^2(T_i).
\end{align}

So, by solving the linear system:
\begin{equation}\label{eq:matrix_solve}
\begin{pmatrix}
    \frac{T_0^2}{T_1^2} & 1 & 0 \\ \frac{T_0^2}{T_2^2} &  0 & 1  \\ 0 & \frac{T_1^3}{T_2^2 T_0} & \frac{T_1}{T_0} \\
\end{pmatrix} 
\begin{pmatrix}
{\sigma}^2_{0}(T_0) \\
{\sigma}^2_{1}(T_0) \\
{\sigma}^2_{2}(T_0)
\end{pmatrix}
=
\begin{pmatrix}
{\sigma}^{\prime 2}_{0,1}(T_0) \\
{\sigma}^{\prime 2}_{0,2}(T_0) \\
{\sigma}^{\prime 2}_{1,2}(T_1)
\end{pmatrix},
\end{equation}
we recover the $\sigma^2_i(T_0)$ needed to compute the entropy rate of the MO-TRNG. Moreover, we note that the coefficients of the matrix  in \Cref{eq:matrix_solve} can be computed from the ratio $T_i/T_j$ which are more stable than the $T_i$ and can be measured precisely using the method from \cite{FL14}.

According to our simulations, \Cref{tab:tab2} shows the expected total jitter volatility per sampling period and the measured total jitter volatility per sampling period using \cite{FL14} for each couple of oscillators $(O_i, O_j)$. Then \Cref{tab:tab1} gives respectively the volatility and measured volatility per period of $O_0$ which are respectively very close to $\sigma_i$ thereby demonstrating the relevance of our method.

\begin{center}
\begin{table}
\resizebox{0.99\textwidth}{!}{
\begin{tabular}{|c|c|c|c|}
\cline{2-4}
\multicolumn{1}{c|}{}  & $(O_0,O_1)$ & $(O_0,O_2)$ & $(O_1, O_2)$ \\ \hline
Expected total jitter & ${\sigma}^{\prime}_{0,1}(T_0)=2.43\, 10^{-3}$ & ${\sigma}^{\prime}_{0,2}(T_0)=3.36\,
    10^{-3}$ & ${\sigma}^{\prime }_{1,2}(T_1)=3.17\, 10^{-3}$ \\ \hline
Measured total jitter & $\hat{\sigma}^{\prime}_{0,1}(T_0)= 2.39\, 10^{-3}$ &   
$\hat{\sigma}^{\prime}_{0,2}(T_0)=3.37\, 10^{-3}$ & $ {\hat{\sigma}^{\prime}}_{1,2}(T_1)=3.19\, 10^{-3}$ \\ \hline
\end{tabular}
}
\caption{Expected total jitter and measured total jitter in our simulations, here $\sigma'_{i,j}(T_k)=\sqrt{T_k \sigma^{\prime 2}_{i,j}}$}\label{tab:tab2}
\end{table}
\end{center}

The code reproducing the experiment can be found in 
the repository \url{https://osf.io/zcj92/?view_only=ded350214f1349b2919e5ef22471c67a}.
\end{example}

\paragraph{General Case}
For a general MO-TRNG, where we have $O_i$ sampled oscillators and $O_0$ the sampling oscillator, we use the preceding trick to obtain  $\sigma_0^2$ and then by performing an internal measurement for the couple of oscillators $(O_0, O_i)$ for $i=1, \ldots, n$, we recover $\sigma_i^2$ from the knowledge of $\sigma_0^2$ using Corollay \ref{cor:transfer}. Thus, by doing $n+1$ measurements, we recover the $n+1$ parameters $\sigma_i^2$ for $i=0, \ldots, n$ which according to the algorithm of \cite{MR4712007} are needed to compute the entropy rate of the MO-TRNG where $O_0$ produces the sampling clock signal. 

More precisely, suppose that a differential measurement method gives us  estimates of transferred jitters ${\hat{\sigma'}}_{i,j}^2 \approx {\sigma'}_{i,j}^2$ for oscillator pairs $(O_i, O_j)$ (for example, one can adjust the method of \cite{FL14} which approximates the accumulated jitter volatility $\sqrt{ T_i {\sigma'}_{i,j}^2 }$). We apply this method for pairs $(i,j)=(0,1),(0,2),(1,2),(0,3),(0,4),\ldots, (0,n-1), (0,n)$ and then retrieve volatilities by solving the resulting system of equations  
\begin{align}\label{eq:volatilities_equations}
  \frac{f_j^2}{f_i^2} \sigma_i^2 +  \sigma_j^2 = {\hat{\sigma'}}_{i,j}^2, 
\end{align}
for $(i,j)=(0,1),(0,2),(1,2),(0,3),(0,4),\ldots, (0,n-1), (0,n)$.
In the matrix form, the equation system becomes
\begin{align*}
\begin{pNiceMatrix} 
\frac{f_1^2}{f_0^2}    & 1 &        & & &   \\ 
\frac{f_2^2}{f_0^2}  &     & 1 & & &   \\         & \frac{f_2^2}{f_1^2} & 1 & & \\
\\
\frac{f_3^2}{f_0^2}  &     & & 1 & & &  \\
\Vdots & & & & \Ddots & &  \\
\frac{f_i^2}{f_0^2}  &     & &  & & 1 \\
\Vdots & & & & & & \Ddots &  \\
\frac{f_{n}^2}{f_0^2}  &     & &  & & & & & 1 \\
\end{pNiceMatrix}
\cdot
\begin{pNiceMatrix}[nullify-dots]
\sigma_0^2 \\
\sigma_1^2 \\
\sigma_2^2 \\
\sigma_3^2 \\
\Vdots \\
\sigma_i^2 \\
\Vdots \\
\sigma_n^2 
\end{pNiceMatrix} = 
\begin{pNiceMatrix}[nullify-dots]
\hat{\sigma'}_{0,1}^2 \\
\hat{\sigma'}_{0,2}^2 \\
\hat{\sigma'}_{1,2}^2 \\
\hat{\sigma'}_{0,3}^2 \\
\Vdots \\
\hat{\sigma'}_{0,i}^2 \\
\Vdots \\
\hat{\sigma'}_{0,n}^2
\end{pNiceMatrix}
\end{align*}
The following result gives the explicit solution and estimates the worst-case relative errors.
\begin{theorem}\label{thm:system_solution}
The explicit solution to \eqref{eq:volatilities_equations} is given by
\begin{align}
\begin{pNiceMatrix}[nullify-dots]
\sigma_0^2 \\
\sigma_1^2 \\
\sigma_2^2 \\
\sigma_3^2 \\
\Vdots \\
\sigma_i^2 \\
\Vdots \\
\sigma_n^2 
\end{pNiceMatrix}
=
   \begin{pNiceMatrix} 
\frac{{f}_{0}^{2}}{2 {f}_{1}^{2}} & \frac{{f}_{0}^{2}}{2 {f}_{2}^{2}} & - \frac{{f}_{0}^{2}}{2 {f}_{2}^{2}}\\
\frac{1}{2} & - \frac{{f}_{1}^{2}}{2 {f}_{2}^{2}} & \frac{{f}_{1}^{2}}{2 {f}_{2}^{2}}\\
- \frac{{f}_{2}^{2}}{2 {f}_{1}^{2}} & \frac{1}{2} & \frac{1}{2} 
\\
-\frac{f_3^2}{2 f_1^2} & -\frac{f_3^2}{2 f_2^2} & \frac{f_3^2}{2 f_2^2} & 1
\\
\Vdots & & & & \Ddots & \\
-\frac{f_n^2}{2 f_1^2} & -\frac{f_n^2}{2 f_2^2} & \frac{f_n^2}{2 f_2^2} &  & & 1 \\
\end{pNiceMatrix}  
\cdot 
\begin{pNiceMatrix}[nullify-dots]
\hat{\sigma'}_{0,1}^2 \\
\hat{\sigma'}_{0,2}^2 \\
\hat{\sigma'}_{1,2}^2 \\
\hat{\sigma'}_{0,3}^2 \\
\Vdots \\
\hat{\sigma'}_{0,i}^2 \\
\Vdots \\
\hat{\sigma'}_{0,n}^2
\end{pNiceMatrix}
\end{align}
and the worst-case relative error is bounded by the condition number, satisfying
\begin{align}
    \kappa_{\infty} \leqslant \left(1+L^2\right)\left(1+\frac{3}{2}L^2\right),
\end{align}
where $L$ denotes the maximum ratio of frequencies 
\begin{align}
L = \max_{i,j} \frac{f_i}{f_j}.
\end{align}
\end{theorem}
Using this result, we obtain concrete error guarantees:
\begin{corollary}[Solutions are stable under comparable frequencies]
If $f_i$ for $i=0,1,\ldots,n$ are of similar magnitude, then $L$ and hence $\kappa_{\infty}$ is a small constant, thus the relative error is well-controlled. In fact, the bound is sharp for $L=1$ when all the frequencies are equal.
\end{corollary}

We note that the system can stated and solved in terms of \emph{accumulated volatilities} $\sigma_i(T_0):=\sqrt{T_0\sigma_i^2}$ and ${\sigma'}_{i,j}(T_i) := \sqrt{T_i {\sigma'}_{i,j}^2 }$, resulting in similar formulas and error bounds.

The implementation can be found in 
the repository \url{https://osf.io/zcj92/?view_only=ded350214f1349b2919e5ef22471c67a}.


\section{Implementations and Discussion}\label{sec:impl}
In the previous section, we presented two methods to compute the jitter of individual ring oscillators using differential measurements:
\begin{itemize}
\item Method 1, which relies on the hypothesis that the variance of the jitter is linear with the period;
\item Method 2, which does not rely on any hypothesis but requires an additional multiplexer or a type-D flip-flop.
\end{itemize}
To compare the two methods, we implemented the schematic shown in Figure \ref{fig:indms} in an Intel Cyclone V FPGA. We added a second flip-flop in each stage to resolve metastability. The output of the second flip-flop $smpl_{ij}$ gives signal $O_j$ sampled by signal $O_i$ (and is equivalent to the output of \Cref{algo:simul} in the simulation).
We tried to challenge the hypothesis in Method 1 by varying the placement and routing of the ring oscillators.

Heve we only present the results of two experiments but we performed other experiments which led to the same conclusion. In the two experiments whose results we describe here, the three oscillators in Figure \ref{fig:indms} are made of $32$ delay elements. In Experiment 1, the frequencies used where: 65.5 MHz for O$_0$, 58.0 MHz for O$_1$ and 70.6 MHz for O$_2$. In Experiment 2, they frequencies were: 65.5 MHz for O$_0$, 58.9 MHz for O$_1$, 71.6 MHz for O$_2$. The difference between the frequencies is due to different choices of placement of the delay elements. 

We acquired 1 Mbits of data for each couple of sampling/sampled oscillators $(O_0, O_1)$, $(O_0, O_2)$, $(O_1, O_2)$ and analysed the output bits using the internal method of \cite{FL14} to recover the variance of the jitter per period of the sampled oscillator. Table \ref{tab:totaljitter} lists the results of the two experiments. Table \ref{tab:method1} (resp. Table \ref{tab:method2}), lists the individual jitter of each ring oscillator we computed using Method 1 (resp. Method 2).

\begin{center}
\begin{table}[h!]
\resizebox{0.99\textwidth}{!}{
\begin{tabular}{|c|c|c|c|}
\cline{2-4}
\multicolumn{1}{c|}{}  & $(O_0,O_1)$ & $(O_0,O_2)$ & $(O_1, O_2)$ \\ \hline
Experiment 1  & ${\sigma'}_{0,1}(T_0)=1.503\, 10^{-3}$ & ${\sigma'}_{0,2}(T_0)=2.532\, 10^{-3}$ &
    ${\sigma'}_{1,2}(T_1)=2.695\, 10^{-3}$ \\ \hline
Experiment 2 & ${{\sigma}'}_{0,1}(T_0)= 1.857\, 10^{-3}$ &   
${{\sigma}'}_{0,2}(T_0)=2.313\, 10^{-3}$ & ${{\sigma}'}_{1,2}(T_1)=3.307\, 10^{-3}$ \\ \hline
\end{tabular}
}
\caption{Measured total jitter of two different implementations in an Intel Cyclone V FPGA, using internal measurement method of \cite{FL14}.}\label{tab:totaljitter}
\end{table}
\end{center}

\begin{center}
\begin{table}[h!]
\begin{tabular}{|c|c|c|c|}
\cline{2-4}
\multicolumn{1}{c|}{}  & $O_0$ & $O_1$ & $O_2$ \\ \hline
Experiment 1  & ${\sigma}_{0}(T_0)=1.305\, 10^{-3}$ & ${\sigma}_{1}(T_0)=1.368\, 10^{-3}$ & ${\sigma}_{2}(T_0)=1.875\, 10^{-3}$ \\ \hline
Experiment 2 & ${{\sigma}}_{0}(T_0)= 1.018\, 10^{-3}$ &   
${{\sigma}}_{1}(T_0)=1.193\, 10^{-3}$ & ${{\sigma}}_{2}(T_0)=2.278\, 10^{-3}$ \\ \hline
\end{tabular}
\caption{Individual jitters, of the same two different implementations in an Intel Cyclone V FPGA, retrieved using Method 1.}\label{tab:method1}
\end{table}
\end{center}

\begin{center}
\begin{table}
\begin{tabular}{|c|c|c|c|}
\cline{2-4}
\multicolumn{1}{c|}{}  & $O_0$ & $O_1$ & $O_2$ \\ \hline
Experiment 1  & ${\sigma}_{0}(T_0)=0.507\, 10^{-3}$ & ${\sigma}_{1}(T_0)=1.801\, 10^{-3}$ & ${\sigma}_{2}(T_0)=2.246\, 10^{-3}$ \\ \hline
Experiment 2 & ${{\sigma}}_{0}(T_0)= 1.164\, 10^{-3}$ &   
${{\sigma}}_{1}(T_0)=1.080\, 10^{-3}$ & ${{\sigma}}_{2}(T_0)=2.195\, 10^{-3}$ \\ \hline
\end{tabular}
\caption{Individual jitters, of the same two different implementations in an Intel Cyclone V FPGA, retrieved using Method 2.}\label{tab:method2}
\end{table}
\end{center}

We see that while the evaluation of the jitter of the individual oscillators mostly agree using the two methods in Experiment 2, there are significant discrepancies in Experiment 1: for instance, with Method 2, the value of $\sigma_0(T_0)$ is half  the value computed Method 1. This shows that the hypothesis that the variance of the jitter per period is linear in the period although reasonable and sometime true is in general at odds with the facts.

\section{Conclusion}\label{sec:conclusion}
The jitter transfer principle is often disregarded in the literature. It is generally used to apply the results obtained by \cite{baudet_security_2011} to compute the entropy rate of a TRNG for the most common designs of oscillator based TRNG where a ring oscillator is sampled by a signal produced by another ring oscillator. With this aim in view, we provided the error bounds needed to ensure the entropy rate is computed with the required precision. But we have also shown that the jitter transfer principle is useful for the measuring the phase noise parameters.
To obtain these parameters, a differential measurement of two oscillators is necessary to filter out global noises and to obtain a good level of accuracy but the outcome of a differential measurement is the total jitter of the oscillators. Using the jitter transfer principle, we have described two methods to obtain individual jitter parameters which is necessary for the computation of the entropy rate of a multi-oscillator ring TRNG following the method of \cite{MR4712007}. A first method uses the hypothesis that the variance of the jitter per period is linear in the period. A second method, does not rely on this assumption but needs a slightly tweaked design of a multi-oscillator based TRNG. Our hardware implementations have demonstrated that the assumption inherent to the first simple method is generally not correct and that the second method is the only reliable way to compute the parameters of the individual jitters of a multi-ring oscillator based TRNG needed to assess its entropy rate. We ran extensive simulations to demonstrate the validity and usefulness of our results and have made our script available. We hope that we have re-established the jitter transfer principle as a cornerstone of ring oscillator based TRNGs and maybe inspire further research on the subject.






\appendix

\section{Proofs}\label{sec:proofs}

\subsection{Proof of \Cref{thm:main}}

Consider the time when the first signal is at its $k$-th rising edge
\begin{align}
    T_k = \inf\{t: \phi_0 + f_0 t +\xi^{0}_t = k\}.
\end{align}
Since $T_k = \inf\{t: f_0 t +\xi^{0}_t = k-\phi_0 \}$, we see that $T_k$ is the moment of hitting $k-\phi_0$ by the stochastic process $f_0 t + \xi_t^{0}$
which is a Brownian motion with drift $f_0$ and volatility $\sigma_0$. The distribution of this hitting time is known~\cite{folks_inverse_1978} to be
\begin{align}
    T_k \sim \mathsf{IG}\left(\frac{k-\phi_0}{f_0},\frac{(k-\phi_0)^2}{\sigma_0^2}\right).
\end{align}
The clock cycles $T_{k+1}-T_k$ are independent (according to the Wiener process properties) and
for the same reasons hitting times~\cite{folks_inverse_1978} are distributed as
\begin{align}\label{eq:clock_cycle}
    T_{k+1}-T_k \sim \mathsf{IG}\left(\frac{1}{f_0},\frac{1}{\sigma_0^2}\right).
\end{align}

The distribution of $\Phi_1(T_k)$, the sampled oscillator phase, is obtained by substituting $T_k$ in the equation of $s_1$. 

The distribution 
$Y=\Phi_1(T_{k+1})-\Phi_1(T_{k})$ is the same as
$\mathsf{N}(f_1 Z, Z \sigma_1^2)$ where $Z = T_{k+1}-T_k$, by the Wiener process properties. Combining this with the distribution of $T_{k+1}-T_k$ derived above, we obtain the claimed variance-mean mixture formula. Since $Y \sim f_1 Z + \sigma_1 \sqrt{Z}\mathsf{N}(0,1)$, we have that 
\begin{align}
    \mathsf{MGF}_Y(s) = \mathbf{E} \exp\left( f_1 s Z + \frac{\sigma_1^2 s^2 Z}{2} \right) = \mathsf{MGF}_Z\left( f_1 s + \frac{\sigma_1^2 s^2}{2}\right).
\end{align}
Using explicit formula for the moment generating function of the inverse Gaussian distribution,
\begin{align}
 \mathsf{MGF}_Y(s) = \exp\left( \frac{f_{0} - \sqrt{f_{0}^{2} - 2 f_{1} s \sigma_{0}^{2} - s^{2} \sigma_{0}^{2} \sigma_{1}^{2}}}{\sigma_{0}^{2}} \right).
\end{align}
Denoting $\alpha = \sqrt{\frac{f_{0}^{2}}{\sigma_{0}^{2} \sigma_{1}^{2}} + \frac{f_{1}^{2}}{\sigma_{1}^{4}}}$, $\beta = \frac{f_{1}}{\sigma_{1}^{2}}$, $\delta = \frac{\sigma_{1}}{\sigma_{0}}$
we reduce this to
\begin{align}
 \mathsf{MGF}_Y(s) = \exp\left( \delta \left(\sqrt{\alpha^{2} - \beta^{2}} - \sqrt{\alpha^{2} - \left(\beta + s\right)^{2}}\right) \right),
\end{align}
which by known formulas \cite{barndorff-nielsen_normal_1982} means that $Y\sim \mathsf{NIG}(\alpha,\beta,0,\delta)$.

The calculations were done with Python library \texttt{SymPy}~\cite{sympy}, as shown in \Cref{lst:mgf_nig}.

\begin{listing}[!ht]
\begin{minted}[
fontsize=\footnotesize,
breaklines
]{python}
import sympy as sm

alpha,beta,f0,f1,s0,s1,phi0,phi1,mu,delta,lmbd,s  = sm.symbols('alpha,beta,f_0 f_1 sigma_0 sigma_1 phi_0 phi_1 mu delta lambda s',positive=True)

# derive as a general mixture
log_MGF_IG = lmbd/mu*(1-sm.sqrt(1-2*mu**2/lmbd*s))
log_MGF_NIG = log_MGF_IG.subs({mu:1/f0,lmbd:1/s0**2})
log_MGF_NIG = log_MGF_NIG.subs({s:f1*s+s1**2*s**2/2})
log_MGF_NIG = log_MGF_NIG.expand().simplify()
log_MGF_NIG

# derive as a normal-inverse gaussian
log_MGF_NIG_2 = delta*(sm.sqrt(alpha**2-beta**2)-sm.sqrt(alpha**2-(s+beta)**2)).simplify()
params = {delta:s1/s0,beta:f1/s1**2,alpha:sm.sqrt((f1*1/s1**2)**2+(f0*1/s0*1/s1)**2)}

# compare
assert (log_MGF_NIG-log_MGF_NIG_2.subs(params)).simplify() == 0
\end{minted}
    \caption{MGF calculations for the NIG distribution.}
    \label{lst:mgf_nig}
\end{listing}

\subsection{Proof of \Cref{cor:mean_variance}}

The result follows the well-known formula for normal variance-mean mixtures~\cite{folks_inverse_1978}, namely
$\mathbf{E}[\mathsf{IG}(\mu,\lambda)]=\mu$ and
$\mathbf{Var}[\mathsf{IG}(\mu,\lambda)] = \frac{\mu^3}{\lambda}$.

\subsection{Proof of \Cref{cor:transfer}}

If $s_0$ was jitter-free, it would have its rising edges at $t_k = \frac{k-\phi_0}{f_0}$ for $k=1,2,\ldots$. \Cref{cor:mean_variance} can thus be rewritten as
\begin{align}
\begin{aligned}
    \mathbf{E}[\Phi_1(T_k)] & = \phi_1 + t_k f_1 \\
    \mathbf{Var}[\Phi_1(T_k)] & = t_k\cdot\left(\left(\frac{f_1}{f_0}\right)^2 \sigma_0^2 + \sigma_1^2\right).
\end{aligned}
\end{align}
We see that the mean and the variance are same as in the oscillator with jitter volatility given by $\sigma'$ such that $
{\sigma'}^2=\left(\frac{f_1}{f_0}\right)^2 \sigma_0^2 + \sigma_1^2$. 
The normal approximation to $\Phi_1(T_k)$ remains to be justified, which can be done with \Cref{thm:main}.

\subsection{Proof of \Cref{thm:normal_approx}}

Let's substitute $h_0 = \sigma_0 / \sqrt{f_0}$, 
$h_1 = \sigma_1 / \sqrt{f_1}$, and $\mu=\frac{f_1}{f_0}$. In these variables the log cumulant function of $Z= \Phi(T_{k+1})-\Phi(T_{k})$ can be expressed as
\begin{align}
    \log\mathbf{E}\exp(s Z) =
    - \frac{\sqrt{1- h_{0}^{2} h_{1}^{2} \mu s^{2} - 2 h_{0}^{2} \mu s }}{h_{0}^{2}} + \frac{1}{h_{0}^{2}}.
\end{align}
Denoting
\begin{align}
x=\mu h_0^2 \left(  h_{1}^{2} s^{2} + 2  s\right),
\end{align}
we can write 
\begin{align}
  \log\mathbf{E}\exp(sZ) & = \frac{x}{2h_0^2}+\frac{x^2}{8 h_0^2} + \frac{x^3 r(x)}{h_0^2} \\
  & = \frac{h_{0}^{2} h_{1}^{4} \mu^{2} s^{4}}{8} + \frac{h_{0}^{2} h_{1}^{2} \mu^{2} s^{3}}{2} + \mu s + s^{2} \left(\frac{h_{0}^{2} \mu^{2}}{2} + \frac{h_{1}^{2} \mu}{2}\right)+ \frac{x^3 r(x)}{h_0^2}
\end{align}
where $r(x)\triangleq \frac{1-\sqrt{1-x}-\frac{x}{2}-\frac{x^2}{8}}{x^3}$ is bounded.

Now, denote
\begin{align}
   v^2 \triangleq \mathbf{Var}[Z] = \mu \left(h_{0}^{2} \mu + h_{1}^{2}\right),
\end{align}
then we have $h_0^2 \leqslant v^2/\mu^2$ and $h_1^2\leqslant v^2/\mu$, and also $|x|\leqslant h_0^2 v^2 s^2 + 2 h_0 v |s|$, thus for $|h_0 v s| < \frac{1}{3}$
\begin{align}
\log\mathbf{E}\exp(sZ) = O(h_0^2 v^4 s^4) + 
O(h_0 v^3 s^3) + \mu s + \frac{v^2 s^2}{2} + O(h_0^4 v^6 s^6) + O(h_0 v^3 s^3).
\end{align}
By changing the variables $s\gets s/v$, we have for $|h_0 s|<\frac{1}{3}$
\begin{align}
   \log\mathbf{E}\exp(s (Z-\mu)/v) = \frac{s^2}{2}+ O(h_0^2 s^4) + O(h_0 s^3) + O(h_0^4 s^6) + O(h_0  s^3). 
\end{align}
Thus with $h_0 \to 0$ we obtain that
\begin{align}
    \frac{Z-\mu}{v} \to \mathsf{N}(0,1),
\end{align}
where the convergence is in the distribution.

\subsection{Proof of \Cref{thm:system_solution}}

Using the matrix identity
\begin{align}
\begin{pNiceMatrix}
A & 0 \\
C & D
\end{pNiceMatrix}^{-1} = 
\begin{pNiceMatrix}
A^{-1} & 0 \\
-D^{-1} C A^{-1} & D^{-1}
\end{pNiceMatrix}
\end{align}
we find that
\begin{align}
M^{-1} =  \begin{pNiceMatrix} 
\frac{{f}_{0}^{2}}{2 {f}_{1}^{2}} & \frac{{f}_{0}^{2}}{2 {f}_{2}^{2}} & - \frac{{f}_{0}^{2}}{2 {f}_{2}^{2}}\\
\frac{1}{2} & - \frac{{f}_{1}^{2}}{2 {f}_{2}^{2}} & \frac{{f}_{1}^{2}}{2 {f}_{2}^{2}}\\
- \frac{{f}_{2}^{2}}{2 {f}_{1}^{2}} & \frac{1}{2} & \frac{1}{2} 
\\
-\frac{f_3^2}{2 f_1^2} & -\frac{f_3^2}{2 f_2^2} & \frac{f_3^2}{2 f_2^2} & 1
\\
\Vdots & & & & \Ddots & \\
-\frac{f_n^2}{2 f_1^2} & -\frac{f_n^2}{2 f_2^2} & \frac{f_n^2}{2 f_2^2} &  & & 1 \\
\end{pNiceMatrix}  
\end{align}
which proves the first part. The bound on the condition number follows by the definition of the condition number 
$\kappa_{\infty} = \| M \|_{\infty} \cdot \| M^{-1} \|_{\infty} $, and the well-known explicit formula for the matrix $p$-norm with $p=\infty$ (the norm equals the maximal absolute row sum) which gives $\|M \|_{\infty} \leqslant 1+L^2$ and $\|M^{-1}\|_{\infty} \leqslant 1+\frac{3}{2}L^2$. When the frequencies are equal, these become equalities so that the bound is attained.

\bibliography{citations}

\newcommand{\etalchar}[1]{$^{#1}$}
\begin{thebibliography}{BGH{\etalchar{+}}23}

\bibitem[Abr74]{abramowitzHandbookMathematicalFunctions1974}
Milton Abramowitz.
\newblock {\em Handbook of {{Mathematical Functions}}, {{With Formulas}}, {{Graphs}}, and {{Mathematical Tables}},}.
\newblock Dover Publications, Inc., USA, May 1974.

\bibitem[BBFV10]{BBFV}
Nathalie Bochard, Florent Bernard, Viktor Fischer, and Boyan Valtchanov.
\newblock True-randomness and pseudo-randomness in ring oscillator-based true random number generators.
\newblock {\em Int. J. Reconfigurable Comput.}, 2010:879281:1--879281:13, 2010.

\bibitem[BGH{\etalchar{+}}23]{bernardLowCostPrecise2023}
Florent Bernard, Arturo Garay, Patrick Haddad, Nathalie Bochard, and Viktor Fischer.
\newblock Low {{Cost}} and {{Precise Jitter Measurement Method}} for {{TRNG Entropy Assessment}}.
\newblock {\em IACR Transactions on Cryptographic Hardware and Embedded Systems}, 2024(1):207--228, December 2023.

\bibitem[BLMT11]{baudet_security_2011}
Mathieu Baudet, David Lubicz, Julien Micolod, and André Tassiaux.
\newblock On the {Security} of {Oscillator}-{Based} {Random} {Number} {Generators}.
\newblock {\em Journal of Cryptology}, 24(2):398--425, April 2011.

\bibitem[BNKS82]{barndorff-nielsen_normal_1982}
O.~Barndorff-Nielsen, J.~Kent, and M.~Sørensen.
\newblock Normal {Variance}-{Mean} {Mixtures} and z {Distributions}.
\newblock {\em International Statistical Review / Revue Internationale de Statistique}, 50(2):145--159, 1982.

\bibitem[Bob16]{bobkov2016proximity}
Sergei~Germanovich Bobkov.
\newblock Proximity of probability distributions in terms of fourier--stieltjes transforms.
\newblock {\em Russian Mathematical Surveys}, 71(6):1021, 2016.

\bibitem[FBB{\etalchar{+}}]{fischer_enhancing_2023}
Viktor Fischer, Florent Bernard, Nathalie Bochard, Quentin Dallison, and Maciej Skórski.
\newblock Enhancing quality and security of the {PLL}-{TRNG}.
\newblock 2023(4):211--237.

\bibitem[FC78]{folks_inverse_1978}
J.~L. Folks and R.~S. Chhikara.
\newblock The {Inverse} {Gaussian} {Distribution} and {Its} {Statistical} {Application}--{A} {Review}.
\newblock {\em Journal of the Royal Statistical Society. Series B (Methodological)}, 40(3):263--289, 1978.

\bibitem[FL14]{FL14}
V.~Fischer and D.~Lubicz.
\newblock {Embedded Evaluation of Randomness in Oscillator Based Elementary TRNG}.
\newblock In L.~Batina and M.~Robshaw, editors, {\em Cryptographic Hardware and Embedded Systems -- CHES 2014}, volume 8731 of {\em LNCS}, pages 527--543. Springer, 2014.

\bibitem[GBF{\etalchar{+}}22]{DBLP:conf/cardis/GarayBFHM22}
Arturo~Mollinedo Garay, Florent Bernard, Viktor Fischer, Patrick Haddad, and Ugo Mureddu.
\newblock An evaluation procedure for comparing clock jitter measurement methods.
\newblock In Ileana Buhan and Tobias Schneider, editors, {\em Smart Card Research and Advanced Applications - 21st International Conference, {CARDIS} 2022, Birmingham, UK, November 7-9, 2022, Revised Selected Papers}, volume 13820 of {\em Lecture Notes in Computer Science}, pages 167--187. Springer, 2022.

\bibitem[HL99]{DLNO}
Ali Hajimiri and Thomas Lee.
\newblock {\em The Design of Low Noise Oscillators}.
\newblock 1999.

\bibitem[KS11]{KS11}
W.~Killmann and W.~Schindler.
\newblock {A proposal for: Functionality classes for random number generators, version 2.0}.
\newblock [online] Available at: \url{ https://www.bsi.bund.de/SharedDocs/Downloads/DE/BSI/Zertifizierung/Interpretationen/AIS_31_Functionality_classes_for_random_number_generators_e.pdf}, 2011.
\newblock Accessed: 2023-10-31.

\bibitem[LF24]{MR4712007}
David Lubicz and Viktor Fischer.
\newblock Entropy computation for oscillator-based physical random number generators.
\newblock {\em J. Cryptology}, 37(2):Paper No. 13, 33, 2024.

\bibitem[MSP{\etalchar{+}}17]{sympy}
Aaron Meurer, Christopher~P. Smith, Mateusz Paprocki, Ond\v{r}ej \v{C}ert\'{i}k, Sergey~B. Kirpichev, Matthew Rocklin, AMiT Kumar, Sergiu Ivanov, Jason~K. Moore, Sartaj Singh, Thilina Rathnayake, Sean Vig, Brian~E. Granger, Richard~P. Muller, Francesco Bonazzi, Harsh Gupta, Shivam Vats, Fredrik Johansson, Fabian Pedregosa, Matthew~J. Curry, Andy~R. Terrel, \v{S}t\v{e}p\'{a}n Rou\v{c}ka, Ashutosh Saboo, Isuru Fernando, Sumith Kulal, Robert Cimrman, and Anthony Scopatz.
\newblock Sympy: symbolic computing in python.
\newblock {\em PeerJ Computer Science}, 3:e103, January 2017.

\bibitem[Wik24]{wikiinverse}
Wikipedia.
\newblock Inverse gaussian distribution --- {W}ikipedia{,} the free encyclopedia, 2024.

\end{thebibliography}

\end{document}